\documentstyle[preprint,pra,aps,epsfig]{revtex}
 
\def\lsim{\mathrel{\rlap{\lower4pt\hbox{\hskip1pt$\sim$}}
    \raise1pt\hbox{$<$}}}         
\def\gsim{\mathrel{\rlap{\lower4pt\hbox{\hskip1pt$\sim$}}
    \raise1pt\hbox{$>$}}}         

\def\be{\begin{equation}}
\def\ee{\end{equation}}
\def\bq{\begin{eqnarray}}
\def\eq{\end{eqnarray}}

\begin{document}
 
\title{  DOUBLE--SPIN TRANSVERSE ASYMMETRIES
\\ IN DRELL--YAN PROCESSES }

\author{ V.~Barone$^{a,b}$,
T.~Calarco$^{c}$ and A.~Drago$^{c}$ \\}

\address{$^{a}$Dipartimento di
Fisica Teorica, Universit\`a di Torino
and INFN, Sezione di
Torino,    10125 Torino, Italy \\
$^{b}$II Facolt{\`a} di Scienze MFN, 15100 Alessandria, Italy\\
$^{c}$Dipartimento di Fisica, Universit{\`a} di Ferrara 
and INFN, Sezione di Ferrara, 44100 Ferrara, Italy  }

\maketitle 
\begin{abstract} 
We calculate the double--spin transverse asymmetries for the 
Drell--Yan lepton pair production in $pp$ and $\bar p p$
collisions. We assume
the transverse and 
the longitudinal polarization densities to be equal at a very small scale, 
as it is suggested by confinement model results.
Using a global fit for the longitudinal distributions,  
we find transverse asymmetries of order 
of $10^{-2}$ at most, in the accessible kinematic regions.
\end{abstract}


\vspace{0.5cm}

\narrowtext

The transverse polarization distribution of quarks (or antiquarks)
$h_1^{q,\bar q}(x,Q^2)$, 
originally introduced by Ralston and Soper \cite{RS}
and studied in more detail in recent years \cite{JJ,AM,CPR}, is totally
unknown from an experimental viewpoint. The reason is that
$h_1$ is a chirally odd quantity \cite{JJ} and hence cannot
be measured in deep inelastic scattering. 
The best way to determine $h_1$ is by $pp$ or, at least
in principle, $\bar p p$ collisions with two transversely polarized
beams \footnote{Also semiinclusive reactions would allow 
extracting $h_1$ but these processes are theoretically 
more complicated 
as they involve combinations of 
twist--3 distributions
and unknown fragmentation functions \cite{JJ2}.}.
The measurement of $h_1$ in proton--proton collisions
is now a chapter of the physics program 
of the STAR and PHENIX experiments
at RHIC \cite{RHIC} and of the proposed HERA-$\vec N$ experiment
at HERA \cite{HERAN}. A
careful analysis of the theoretical situation 
is therefore called for. 

What is planned to be measured is the double--spin transverse
asymmetry, whose operational definition is
\be
A_{TT} = 
\frac{d \sigma_{\uparrow \uparrow} - 
d \sigma_{\uparrow \downarrow}}{d \sigma_{\uparrow \uparrow} +
d \sigma_{\uparrow \downarrow}}\,,
\label{1}
\ee
where the arrows denote the transverse directions along which 
the two colliding hadrons are polarized. 

Among the 
processes initiated by $pp$ scattering one can make a 
selection choosing those which are expected to yield the 
largest $A_{TT}$. 
In fact, since there is no analogue of $h_1$ for gluons \cite{JJ}, 
all processes  taking place at the partonic level
via $qg$ or $gg$ scattering give a large contribution
to the denominator of (\ref{1}) and a vanishing one 
to the numerator, thus producing a negligibly small $A_{TT}$ 
\cite{Ji,JS}. 
This leaves us with only one promising reaction: the Drell-Yan
lepton pair production. The double--spin transverse asymmetry for this
process was 
calculated in \cite{Ji,BS,BS2} and found to be relatively large
($\sim 0.1-0.3$ at $\sqrt{s} = 100$ GeV and for a dilepton mass 
$M = 10$ GeV). The 
basic assumption made in Refs.~\cite{Ji,BS} in order to calculate
$A_{TT}$ was that $h_1^q$ and $h_1^{\bar q}$ are equal 
to the helicity distributions $\Delta q$ and $\Delta \bar q$, 
respectively, 
at the experimentally probed momentum scales. Now, since it is
known from confinement model calculations \cite{BCD} that
$h_1^q \simeq \Delta q$ and $h_1^{\bar q} \simeq \Delta \bar q$
only at very small $Q^2$ ($Q^2 \lsim 0.5$ GeV$^2$), the assumption 
above amounts to neglecting the difference between the 
QCD evolution of $h_1^{q, \bar q}$ and that of $\Delta q, \Delta
\bar q$ \footnote{A different procedure to relate the transverse
distribution to the longitudinal and unpolarized distributions is adopted
in \cite{BS2} but again the peculiarity of the QCD evolution
of $h_1$ is overlooked and results quantitatively 
similar to those of \cite{Ji,BS} are obtained.}. 
Now, the evolution of $h_1$ is driven at leading 
order by gluon emission 
and the first anomalous dimension for this process, although
non vanishing, is rather small \cite{AM}, hence
not much different from the corresponding anomalous dimension 
of the helicity distributions, which vanishes exactly. 
However, if one goes from the 
space of moments to the $x$--space, the difference between 
the evolutions of $h_1$ and $\Delta q$ becomes 
large \cite{BCD,B}  at small $x$ and affects all the 
observables which are sensitive to this kinematic
region. 
This was first pointed out in \cite{BCD}, where 
a confinement model calculation of $h_1$ and $A_{TT}$ was 
presented. It was found there that $A_{TT}$ 
is rather small, of order of few percent. 
In what follows we shall confirm and 
extend the findings 
of \cite{BCD} about $A_{TT}$ in a more model independent way.

Considering only the quark--antiquark annihilation 
graph\footnote{The contribution of next--to--leading order (NLO) diagrams has
been crudely estimated in \cite{MS}
and found to correct 
the dominant term in the asymmetries by less than 10 \%. 
However a complete computation of the NLO diagrams and 
of the NLO splitting functions for $h_1$ is not available yet, 
hence the result of \cite{MS} can be taken only as an evaluation
of the theoretical uncertainty of the leading order result.}
the double--spin transverse asymmetry for the $pp$ (or $\bar p p$) 
Drell--Yan 
process mediated by a virtual photon is given by
\begin{equation}
A_{TT}=a_{TT}\,
\frac{\sum_q e_q^2 h_1^q(x_a, M^2) h_1^{\bar q}(x_b,M^2) +
( a \leftrightarrow b)}{\sum_q e_q^2 q(x_a,M^2) \bar q(x_b,M^2)
+ ( a \leftrightarrow b) }\, ,
\label{2}
\end{equation}
where we have labeled by $a,b$ the two incoming hadrons,
the virtuality $M^2$ of the quark and antiquark distributions
is the squared mass of the produced dilepton pair, and $x_a,x_b,M^2$
are related to the center of mass energy $\sqrt{s}$ by $x_a x_b =
\frac{M^2}{s}$.
The partonic asymmetry $a_{TT}$ is calculable in perturbative QCD
\cite{RS} and in the dilepton center-of-mass frame reads
\be
a_{TT} = \frac{\sin^2 \theta \, \cos 2\phi}{1 + \cos^2 \theta}\,,
\label{3}
\ee
where $\theta$ and $\phi$ are the polar and azimuthal angles
of the lepton momentum with respect to the beam and the 
polarization axes, respectively. The value of $a_{TT}$
varies 
between $-1$ and $1$. 

Another potentially interesting Drell--Yan process is the one mediated by a 
$Z^0$, for which the transverse double--spin asymmetry reads \cite{BS}
\be
A_{TT}^{Z}=a_{TT}\,
\frac{\sum_q (V_q^2 - A_q^2)\, h_1^q(x_a, M^2) h_1^{\bar q}(x_b,M^2) +
( a \leftrightarrow b)}{\sum_q (V_q^2 + A_q^2) \,
q(x_a,M^2) \bar q(x_b,M^2)
+ ( a \leftrightarrow b) }\, ,
\label{4}
\ee
having denoted by $V_q$ ($A_q$) the vector (axial) coupling 
of the flavor $q$ to the $Z^0$ (see for instance, \cite{LP}).

Let us come to our calculation. 
The only result we borrow from model computations are the 
approximate equalities 
\be
h_1^q (x,Q_0^2) \simeq \Delta q(x, Q_0^2)\,,\;\;\;\;
h_1^{\bar q} (x,Q_0^2) \simeq \Delta {\bar q}(x, Q_0^2)
\label{1b}
\ee
 at a
scale $Q_0^2 \lsim 0.5$ GeV$^2$. This is the starting point of the 
calculation. Instead of resorting
to models,  we use a global
parametrization for the parton distributions. 
Since the equalities (\ref{1b}) are 
expected to be valid only at very small momentum transfer, 
we adopt the GRV \cite{GRV} leading order fit which provides
the input polarized densities $\Delta q, \Delta \bar q$ 
at $Q_0^2 = 0.23$ GeV$^2$. 
Then the transverse distributions are evolved up to 
$M^2$ by solving
the appropriate Altarelli--Parisi equation with the 
leading order splitting
function computed in \cite{AM}. 

For illustration
we show in  Fig.~1  the input and the 
evolved distributions for the dominant $u,\bar u$ flavor  
(the situation is similar 
 for the other flavors but  we omit their plots for brevity). The difference 
between $h_1^u$ and $\Delta u$ at the scale $Q^2 = 25$ GeV$^2$
in the low--$x$ region is evident: $h_1^u$ is smaller than 
$\Delta u$ for $x \simeq 10^{-2}$, whereas 
$h_1^{\bar u}$ is larger in absolute value than its longitudinal 
counterpart  in the intermediate--$x$
region and smaller at low $x$. This behavior of the 
parton distributions is responsible for
the difference between the longitudinal and the transverse
double--spin asymmetries, as we shall see below.

We present now  the results for $A_{TT}$. 
 Fig.~2 displays $\vert A_{TT}/a_{TT} \vert$ as a function of 
$M^2$ for $pp$ collisions with two different c.m. energies. 
Note that the sign of $A_{TT}/a_{TT}$ is negative.
In Fig.~3 $\vert A_{TT}/a_{TT}\vert$ is presented as a function 
of the Feynman variable $x_a-x_b$ for two $M^2$ values. 
In the kinematic range covered by the figures, which 
 roughly corresponds to the  experimentally accessible region, 
the transverse asymmetry
is at most few percent. It is also 
 a decreasing function of 
$\sqrt{s}$ for fixed $M^2$, and an increasing function of $M^2$
for fixed $\sqrt{s}$.
Therefore, reaching high values of $\sqrt{s}$, such as those at which RHIC
will operate, does not help getting 
a sizable asymmetry. On the other hand, 
since  $M^2$ and $\sqrt{s}$ are experimentally correlated, the relatively
small c.m. energy of HERA-$\vec N$ ($\sqrt{s} 
\simeq 50$ GeV) allows exploring only the low--mass spectrum of 
dileptons ($M^2 \simeq 20$ GeV$^2$), where again the transverse asymmetry 
is expected to be small. Hence the conclusion of a transverse
asymmetry of order of $10^{-2}$ in the  experimentally relevant
kinematic region seems to be inescapable.

For comparison, in Figs.~2,3 we also display the
double--spin longitudinal asymmetry $\vert A_{LL}/a_{LL} \vert$ 
obtained similarly from  the GRV parton distributions.
In the kinematic region shown in the figures
the longitudinal asymmetry is larger than the transverse one 
as a consequence of the low--$x$ behavior of the 
corresponding polarization distributions.

Analogous curves for the transverse asymmetry in 
the $\bar p p$ Drell--Yan process are plotted in Figs.~4,5. 
As we expected, the asymmetries are systematically  larger 
than in the $pp$ case, 
but the exploration of this process 
seems to be beyond the present experimental 
possibilities.

In Fig.~6  the results for  
the Drell--Yan production via $Z^0$ exchange 
are presented. In this case, 
for kinematic reasons (remember that $x_a x_b =\frac{M^2}{s}$), 
 the asymmetries reflect 
the behavior of the 
distributions in the intermediate--$x$ region, where the 
transverse densities are larger in absolute value than the 
longitudinal ones. Thus $\vert A_{TT}^Z/a_{TT} \vert$ 
lies above its longitudinal counterpart, although it remains 
of order of few percent.

In conclusion, we presented a calculation of the double-spin
transverse asymmetries in various Drell--Yan processes, 
based on the assumption (\ref{1b}) and on a global fit 
for the input distributions. 
The two novelties of our approach, which represent
an improvement with respect to previous computations, are: 
{\it i)} the equality (\ref{1b}) is strictly enforced only
at a very small scale, that is the 
typical scale at which quark models
describe the nucleon; {\it ii)} the $Q^2$ evolution  
of $h_1$ is properly treated. 
The outcome of our calculation is similar to a
previous confinement model result \cite{BCD}, namely the 
transverse asymmetries are rather small, of order of 
$10^{-2}$, in the accessible kinematic regions. 
An interesting finding is that $\vert A_{TT} \vert$ 
decreases with increasing center of mass energy. 
In order to maximize the asymmetry a delicate balance 
between $\sqrt{s}$ and the dilepton invariant mass $M^2$
should be found, although it appears unlikely to get 
values larger than few percent. As for the comparison 
with the longitudinal asymmetry, it is 
interesting to notice that we expect to have 
$\vert A_{TT}^Z\vert $ larger than $\vert A_{LL}^Z\vert $
in the $Z^0$--mediated Drell--Yan process. 
All these predictions will hopefully 
be tested by future experiments.

\pagebreak

\begin{center}
{\Large \bf Figure Captions}

\end{center}

\begin{itemize}

\item[Fig.~1]
The longitudinal and transverse polarization distributions
for $u$ and $\bar u$ at the input scale $Q_0^2 = 0.23$ GeV$^2$
and evolved up to $Q^2 = 25 $GeV$^2$.

\item[Fig.~2]
The Drell-Yan double-spin 
transverse asymmetry
$\vert A_{TT}/a_{TT} \vert$ for $pp$ collisions as a function 
of $x_a-x_b$ (dot-dashed line: $M^2=100$ GeV$^2$;  
 solid line: $M^2=25$ GeV$^2$). For comparison, 
the double
longitudinal asymmetry  $\vert A_{LL}/a_{LL}\vert$ is shown 
for $M^2= 25$ GeV$^2$ (dashed line). All curves are obtained with
$\sqrt{s}=100$ GeV.

\item[Fig.~3]
Dependence on $M^2$ of the Drell--Yan  double-spin transverse 
asymmetry for $pp$ at 
$x_a-x_b=0$ (dot-dashed line: $\sqrt{s}=100$ GeV;  
 solid line: $\sqrt{s}=500$ GeV). The longitudinal counterpart is also 
plotted (dashed curve) for $\sqrt{s}=500$ GeV. 

\item[Fig.~4]
Same as Fig.~2, for $\bar p p$ collisions.

\item[Fig.~5]
Same as Fig.~3, for $\bar p p$ collisions.

\item[Fig.~6]
The double--spin asymmetry for the $Z^0$--mediated Drell--Yan 
process as a function of $x_a-x_b$ for $\sqrt{s} = 500$ GeV
(solid line: the transverse asymmetry $\vert A_{TT}^Z/a_{TT} \vert$; 
dashed line: the longitudinal asymmetry $\vert A_{LL}^Z/a_{LL} \vert$).

\end{itemize}

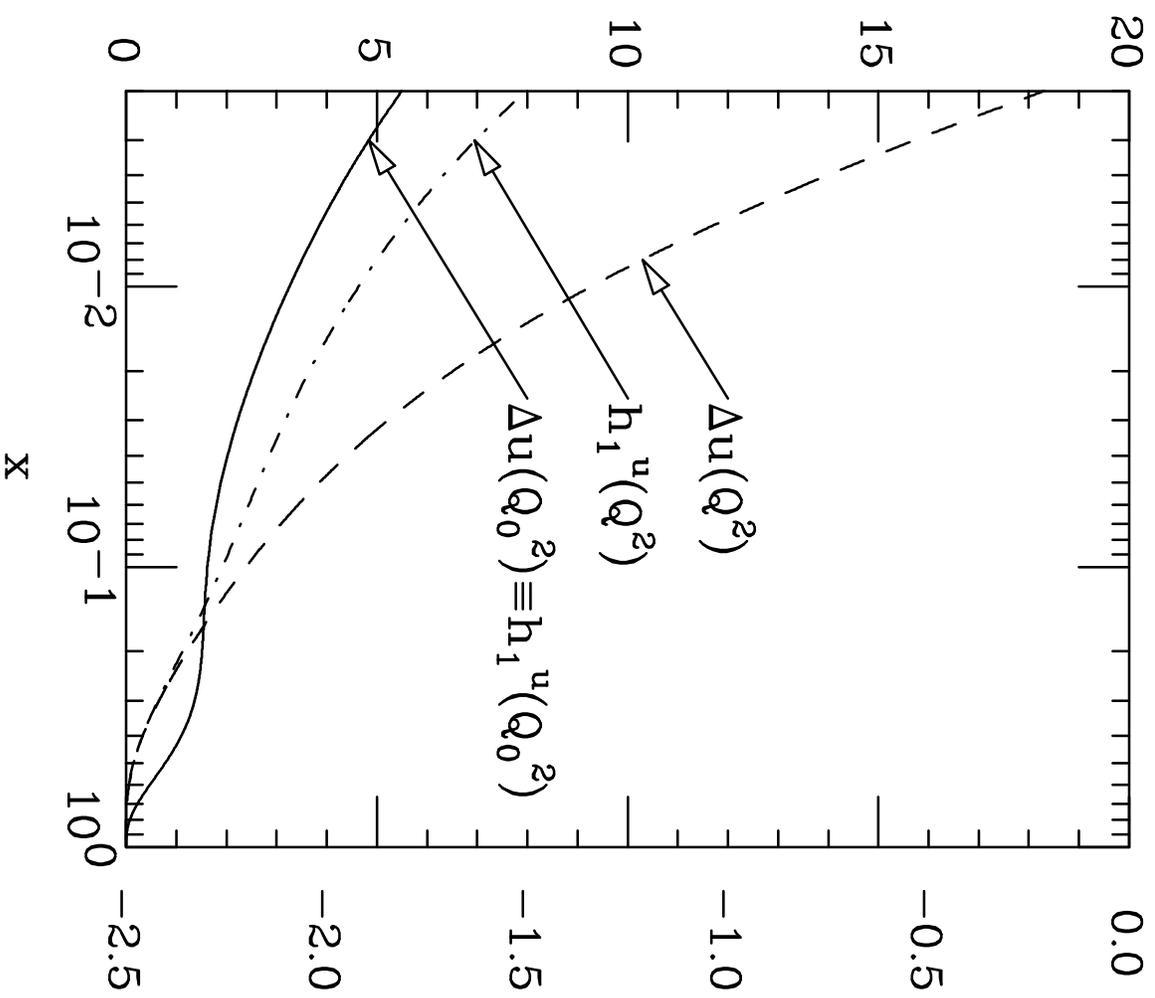
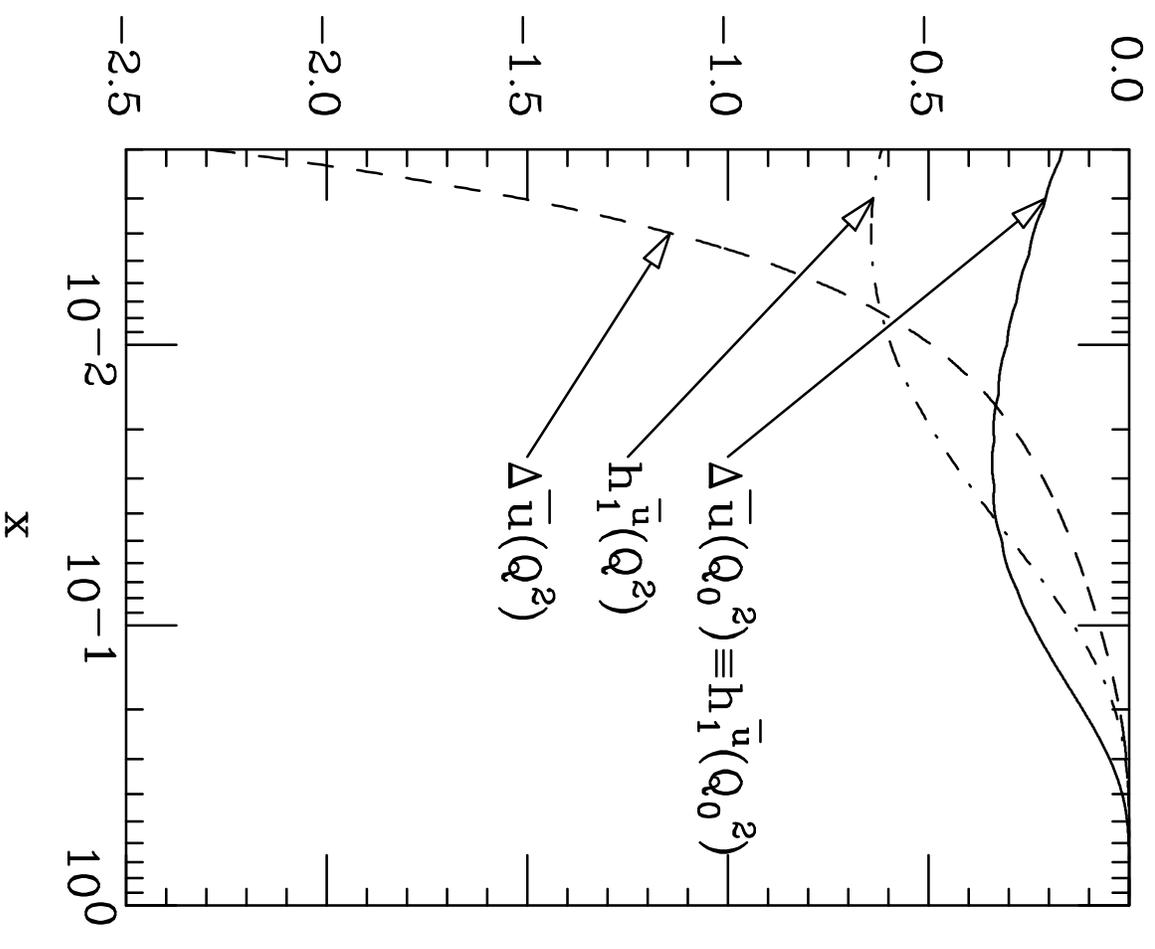
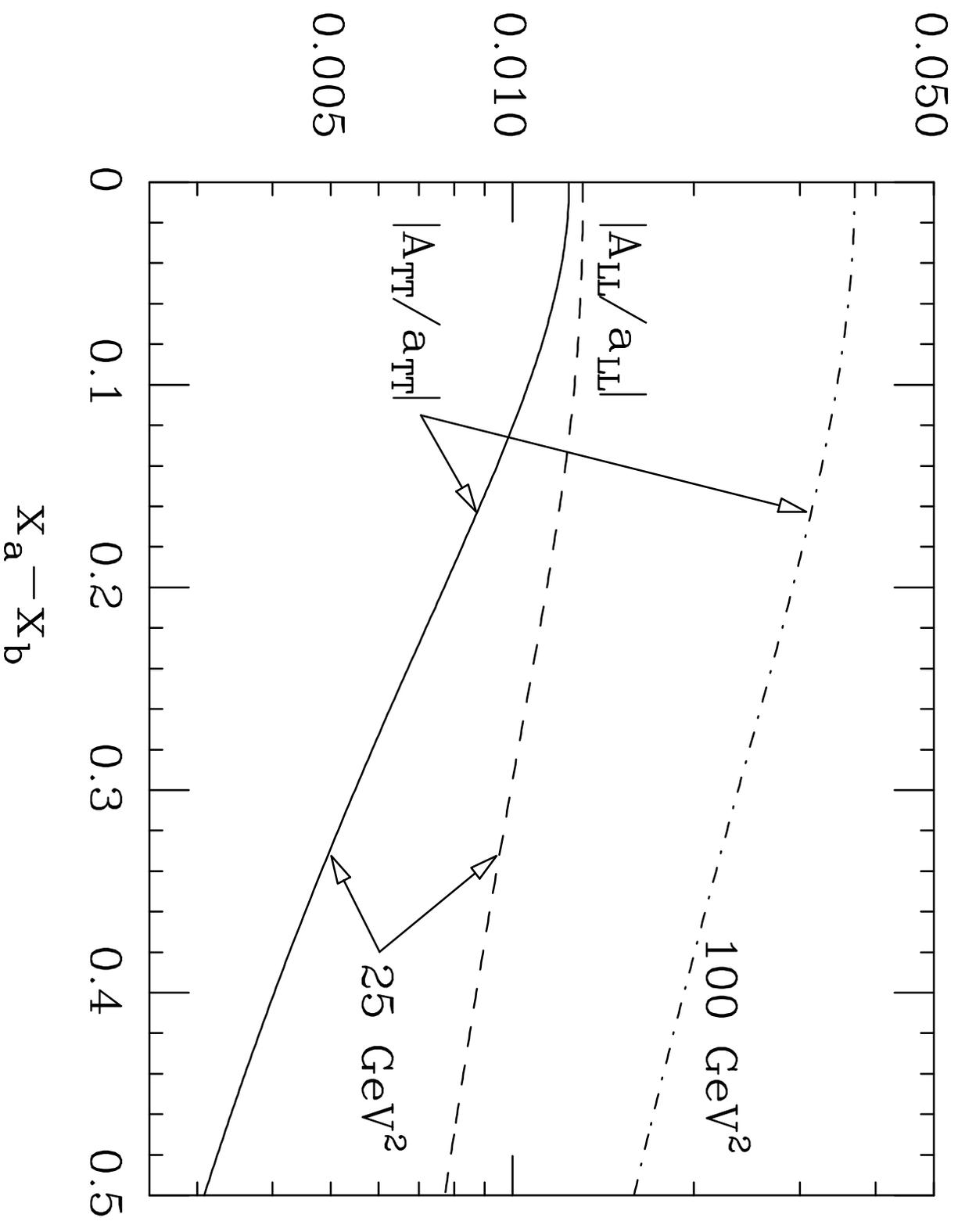
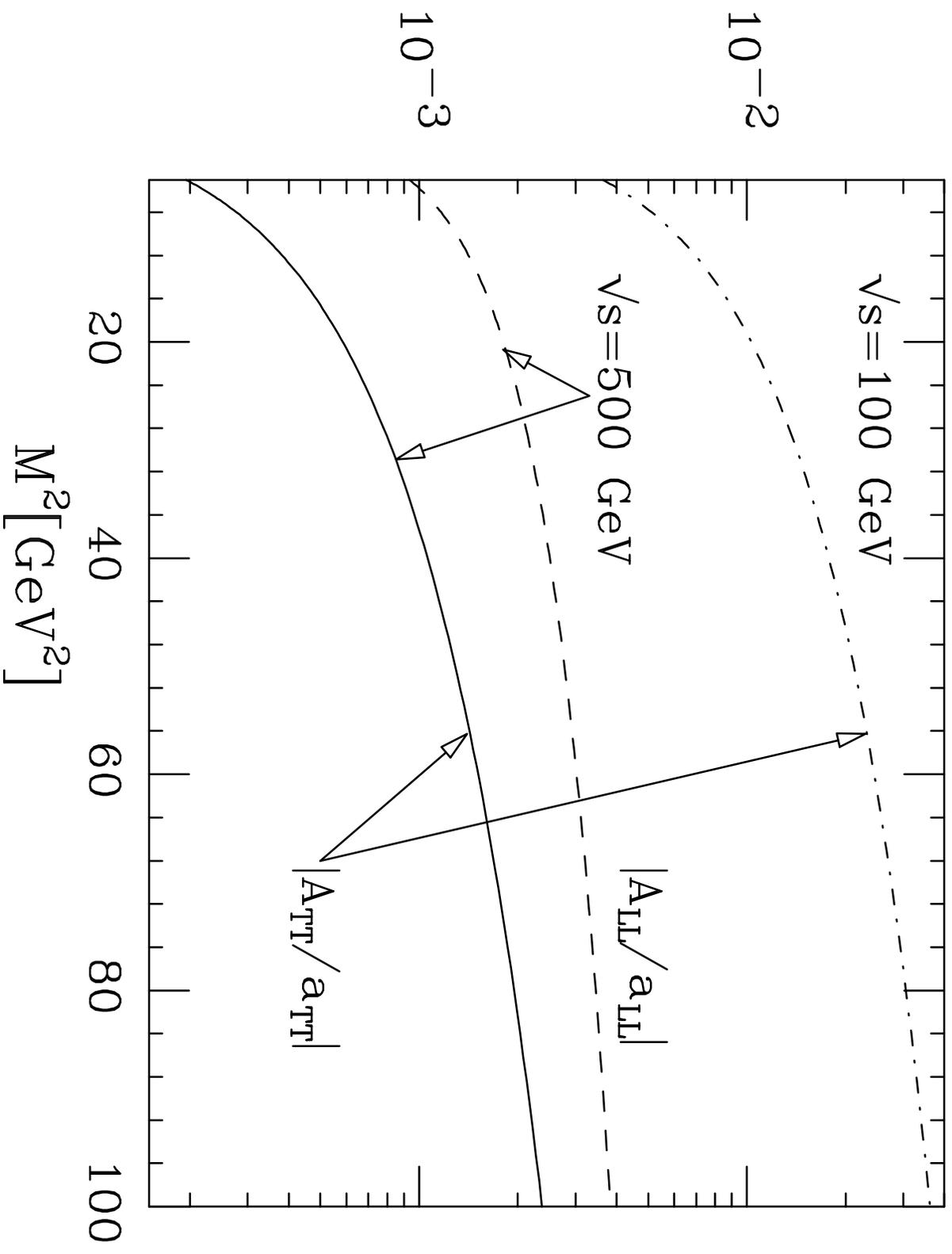
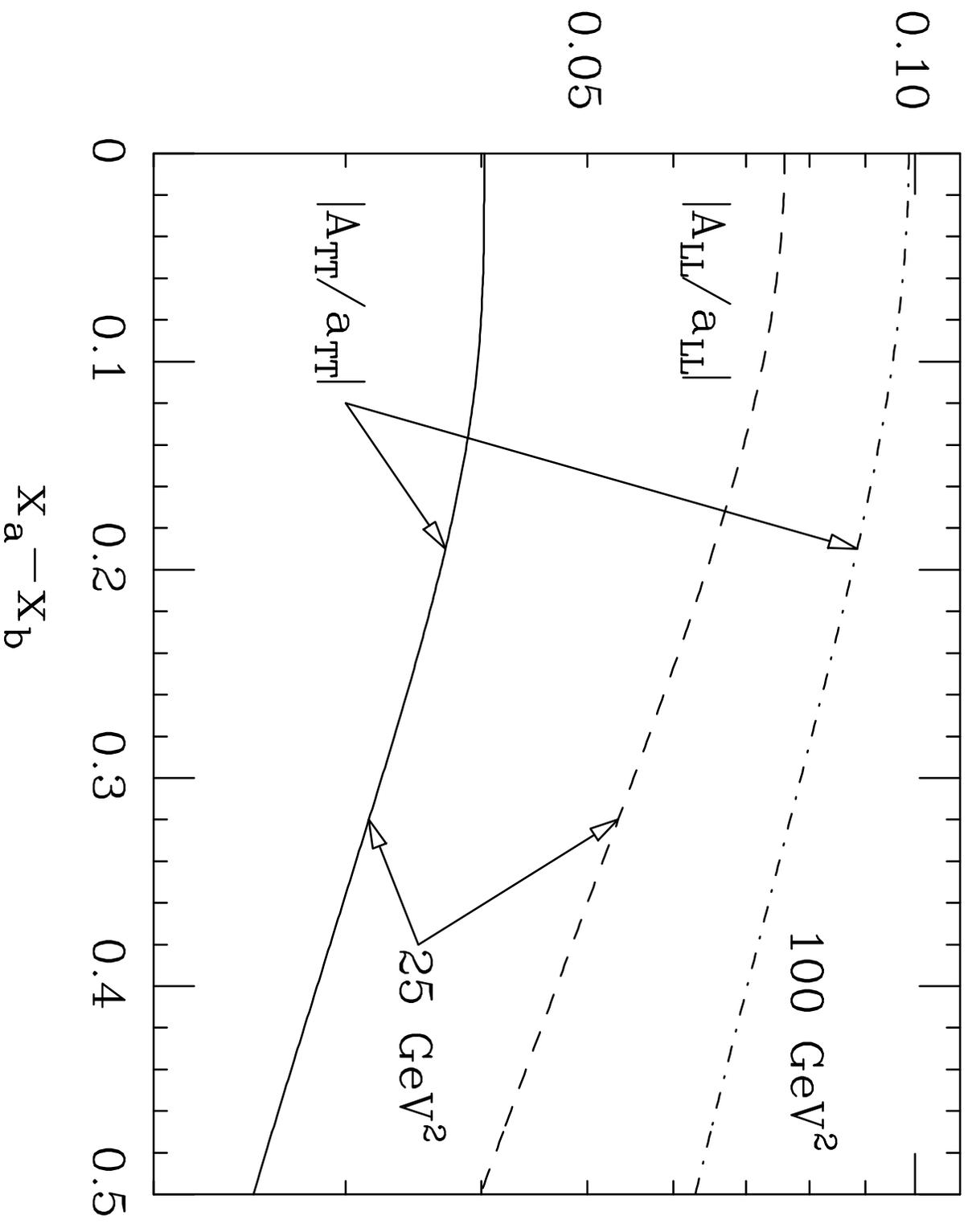
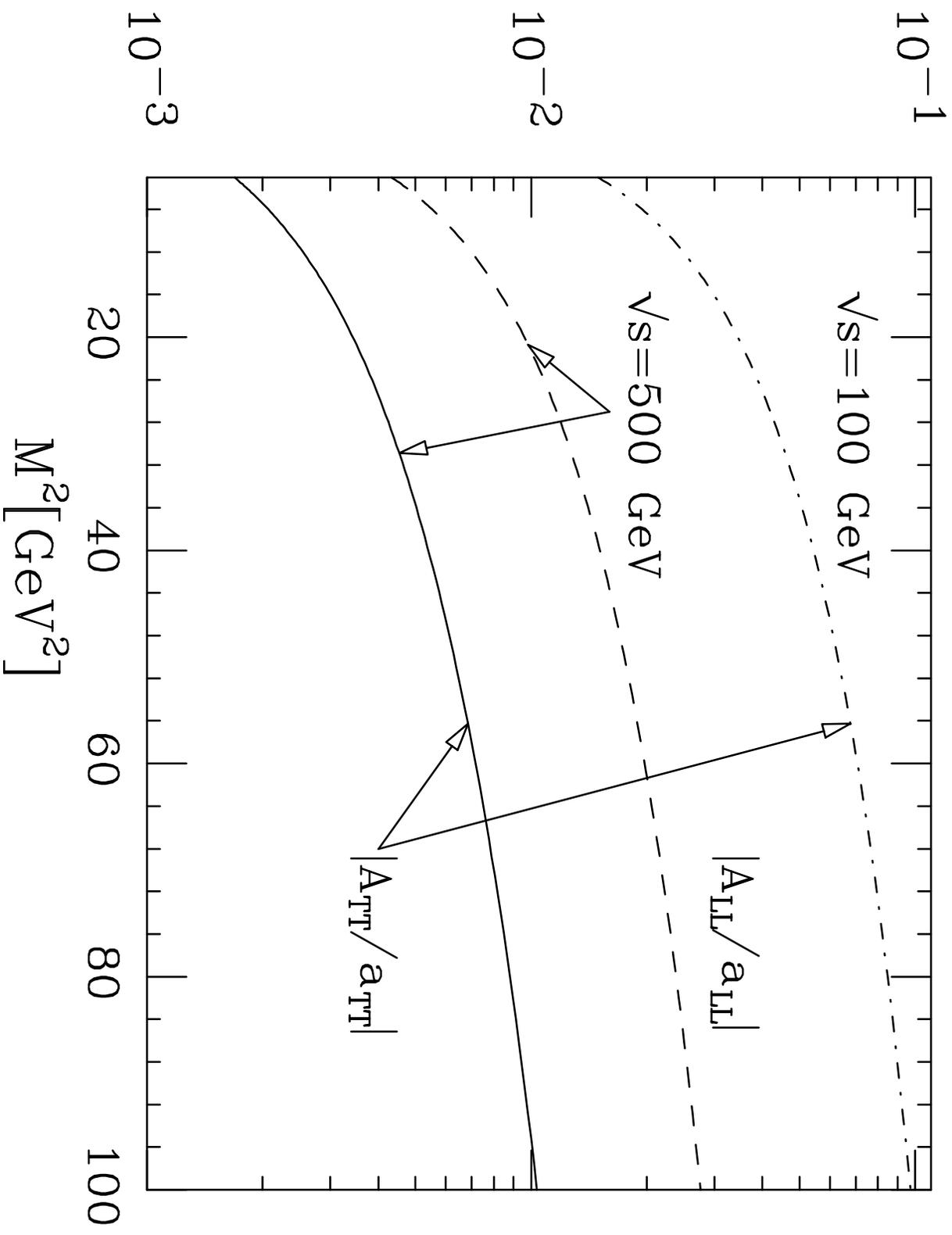
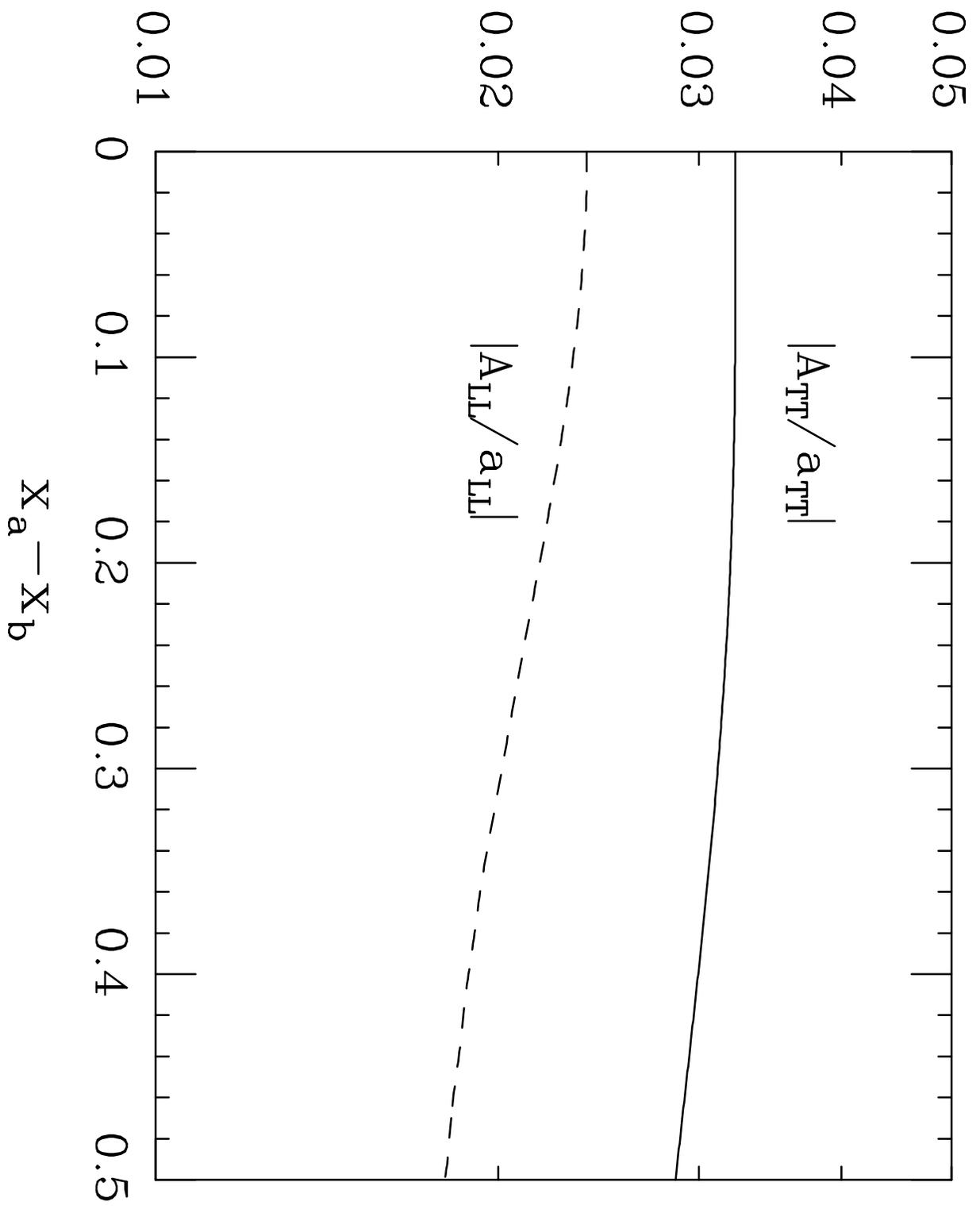
\end{document}